\documentclass[12pt]{iopart}
\usepackage{iopams}
\usepackage{graphicx}

\newcommand{\dst}{\displaystyle}

\newcommand{\rhm}{\rho_{m}}
\newcommand{\prm}{p_{m}}

\newcommand{\pat}{\partial}

\newcommand{\No}{N}

\newcommand{\tilt}{\tilde{t}}
\newcommand{\tilrho}{\tilde{\rho}}
\newcommand{\tilp}{\tilde{p}}
\newcommand{\tilm}{\tilde{m}}
\newcommand{\tilR}{\tilde{R}}
\newcommand{\tilH}{\tilde{H}}
\newcommand{\tilV}{\tilde{V}}

\newcommand{\tilZ}{\tilde{Z}}
\newcommand{\tilphi}{\tilde{\phi}}
\newcommand{\tilrhm}{\tilde{\rho}_m}
\newcommand{\tilprm}{\tilde{p}_m}
\newcommand{\tilrhmz}{\tilde{\rho}_{m0}}

\newcommand{\const}{\mathrm{const}}

\begin{document}
\title{Analysis of inflationary cosmological models in gauge theories of gravitation}
\author{A V Minkevich$^{1,2}$ and A S Garkun$^1$}
\address{$^1$ Department of Theoretical Physics, Belarussian State University, F. Skoriny av. 4,
Minsk 20050, Belarus}
\address{$^2$  Department of Physics and Computer Methods, Warmia
and Mazury University in Olsztyn, Poland}
\eads{\mailto{minkav@bsu.by}, \mailto{garkun@bsu.by}}
\begin{abstract}
Inflationary homogeneous isotropic cosmological models filled by scalar fields and
ultrarelativistic matter are examined in the framework of gauge theories of gravitation. By using
quadratic scalar field potential numerical analysis of flat, open and closed models is curried out.
Properties of cosmological models are investigated in dependence on indefinite parameter of
cosmological equations and initial conditions at a bounce. Fulfilled analysis demonstrates regular
character of all cosmological models.
\end{abstract}

\section{Introduction}

As it was shown in a number of papers (see \cite{mc1,mc2} and Refs. given therein), general feature
of homogeneous isotropic cosmological models built in the framework of gauge theories of
gravitation (GTG) \cite{mca3,mca4} is their regular bouncing character. All cosmological solutions
of generalized cosmological Friedmann equations (GCFE) deduced for homogeneous isotropic models in
GTG are regular in metrics, Hubble parameter and its time derivative because of gravitational
repulsion effect at extreme conditions (extremely high energy densities and pressures), which takes
place in the case of usual gravitating systems satisfying energy dominance condition \cite{mc3}. In
the case of inflationary cosmological models filled by interacting scalar field and usual
gravitating matter, indicated above property of cosmological solutions was obtained without
concretization explicit form of scalar field potential and equation of state of gravitating matter
\cite{mc1}. However, we have to use their explicit form to obtain and analyze cosmological
solutions of GCFE.

The present paper is devoted to analysis of inflationary cosmological models filled by
noninteracting scalar fields and ultrarelativistic matter. In Sec.~2 the GCFE for such inflationary
cosmological models are introduced. In Sec.~3 some general properties of its solutions are
examined. In Sec.~4 and 5 by using quadratic scalar field potential numerical analysis of
cosmological solutions for flat, open and closed models is carried out.

\section{Generalized cosmological Friedmann equations in GTG}

The GCFE deduced in the frame of Poincare GTG and metric-affine GTG have the following form [6-8]
\begin{eqnarray}
&\displaystyle{\frac{k}{R^2}+\left\{\frac{d}{dt}\ln\left[R\sqrt{\left|1+\alpha\left(\rho-
3p\right)\right|}\,\right]\right\}^2=\frac{8\pi G}{3}\,\frac{\rho+
\frac{\alpha}{4}\left(\rho-3p\right)^2}{1+\alpha\left(\rho-3p\right)}
\, ,}\label{e1}\\
&\displaystyle{\frac{\left[\dot{R}+R\left(\ln\sqrt{\left|1+\alpha\left(\rho-
3p\right)\right|}\,\right)^{\cdot}\right]^\cdot}{R}= -
 \frac{4\pi G}{3}\,\frac{\rho+3p-\frac{\alpha}{2}\left(\rho-3p\right)^2}{
1+\alpha\left(\rho-3p\right)}\, ,}\label{e2}
\end{eqnarray}
where $R(t)$ is the scale factor of Robertson-Walker metrics, $\rho$ is energy density, $p$ is
pressure, $k=+1,0,-1$ for closed, flat, open models respectively, $G$ is Newton's gravitational
constant (the light velocity $c=1$), $\alpha$ is indefinite parameter with inverse dimension of
energy density, a dot denotes differentiation with respect to time $t$.  Together with Newton's
gravitational constant $G$ the parameter $\alpha$ is fundamental physical constant of relativistic
isotropic cosmology built in the frame of GTG. The value of $\alpha^{-1}$ determines the scale of
extremely high energy densities \cite{mc3}, which will be supposed to be less than the Planckian
one. From equations (1)--(2) follows the energy conservation law in usual form
\begin{equation}
\label{e3}
\dot{\rho}+3H\left(\rho+p\right)=0 \qquad \left(H=\frac{\dot{R}}{R}\right).
\end{equation}
Obviously, the solutions of GCFE (\ref{e1})--(\ref{e2}) depend on values of $G$ and $\alpha$.
However, we can exclude them and transform equations (\ref{e1})--(\ref{e2}) to dimensionless form
by means of the following transformation of the time $t$ and functions $R(t)$, $\rho(t)$, $p(t)$
\begin{equation}
\label{e4}
\begin{array}{lcl}
t\to\tilt=t\sqrt{\frac{G}{\alpha}},&\qquad\qquad & R\to\tilR=R\sqrt{\frac{G}{\alpha}},\\
\rho\to\tilrho=\alpha\,\rho,& & p\to\tilp=\alpha\,p
\end{array}
\end{equation}
As result we obtain
\begin{eqnarray}
&\displaystyle{\frac{k}{\tilR^2}+\left[\frac{d}{d\tilt}\ln\left(\tilR\sqrt{\left|1+\tilrho-
3\tilp\right|}\,\right)\right]^2=\frac{8\pi}{3}\,\frac{\tilrho+
\frac{1}{4}\left(\tilrho-3\tilp\right)^2}{1+\tilrho-3\tilp}
\, ,}\\
&\displaystyle{\frac{\left[\tilR'+\tilR\left(\ln\sqrt{\left|1+\tilrho-
3\tilp\right|}\,\right)'\right]'}{\tilR}=
-\frac{4\pi}{3}\,\frac{\tilrho+3\tilp-\frac{1}{2}\left(\tilrho-3\tilp\right)^2}{
1+\tilrho-3\tilp}\, ,}
\end{eqnarray}
where a prime denotes the differentiation with respect to $\tilt$. The form of equation (\ref{e3})
does not change by transformations (\ref{e4})
\begin{equation}
\label{e7}
\tilrho'+3\tilH\left(\tilrho+\tilp\right)=0 \qquad
\left(\tilH=\frac{\tilR'}{\tilR},\quad \tilH=H\sqrt{\frac{\alpha}{G}}\right).
\end{equation}

In the case of models filled by scalar field $\phi$ with potential $V=V(\phi)$ and gravitating
matter with energy density $\rhm$ and pressure $\prm$ we have
\begin{equation}
\label{e8}
\rho=\frac{1}{2}\dot{\phi}^2+V+\rhm, \qquad p=\frac{1}{2}\dot{\phi}^2-V+\prm.
\end{equation}
By using the transformation (\ref{e4}) we obtain from (\ref{e8}) the following dimensionless
expressions:
\begin{equation}
\tilrho=\frac{1}{2} \tilphi'^2+\tilV+\tilrhm, \qquad \tilp=\frac{1}{2}
\tilphi'^2-\tilV+\tilprm,
\end{equation}
where $\tilphi=\phi\sqrt{G}$, $\tilV=\alpha\,V$, $\tilrhm=\alpha\,\rhm$, $\tilprm=\alpha\,\prm$. By
neglecting the interaction between scalar field and gravitating matter, from conservation law
(\ref{e7}) follows the equation for scalar field
\begin{equation}
\tilphi''+3\tilH\tilphi'=-\frac{\pat\tilV}{\pat\tilphi}
\end{equation}
and conservation law for gravitating matter
\begin{equation}
\tilrhm'+3\tilH(\tilrhm+\tilprm)=0.
\end{equation}

To integrate equations (5)--(6) with $\tilrho$ and $\tilp$ defined by (9) we have to know the
equation of state of gravitating matter $\prm=\prm(\rhm)$. To build realistic cosmological models
we have to take into account the change of equation of state under their evolution in accordance
with properties of gravitating matter, but it is generally a complicated physical problem. To study
cosmological models near a bounce and simplify further analysis we will use below the equation of
state for ultrarelativistic matter $\prm=\frac{1}{3}\rhm$.

In considered case from equation (11) follows the integral for gravitating matter
\begin{equation}
\tilrho \tilR^{4}=\const .
\end{equation}
Then the GCFE (5)--(6) can be transformed in accordance with Ref.~\cite{mc3} to the following form:
\begin{eqnarray}
\label{e9}\fl
 \left[
 \tilH\left(
   \tilZ+3\tilphi'^2
   \right)
 +3\frac{\pat\tilV}{\pat\tilphi}\tilphi'
 \right]^2
+\frac{k}{\tilR^2}\,\tilZ^2 
=\frac{8\pi}{3}\,
 \left[
   \tilrhm+\frac{1}{2}\tilphi'^2+\tilV +\frac{1}{4}\,
   \left(4\tilV-\tilphi'^2\right)^2
 \right]
\,\tilZ,
\end{eqnarray}
\begin{eqnarray}
\label{e10}\fl
\tilH'\left(
    \tilZ+3\tilphi'^2
\right)
+3\tilH^2 \left(\tilZ-\tilphi'^2\right)
&+3\left[
    \frac{\pat^2\tilV}{\pat\tilphi^2}\tilphi'^2-\left(\frac{\pat\tilV}{\pat\tilphi}\right)^2
\right]
\nonumber\\
&=8\pi\left[
    \tilV+\frac{1}{3}\tilrhm
    +\frac{1}{4}\left(4\tilV-\tilphi^2\right)^2
\right]-\frac{2k}{\tilR^2}\tilZ,
\end{eqnarray}
 where $\tilZ=1+4\tilV-\tilphi'^2$.

The greatest part of solutions of GCFE (13)--(14) are inflationary cosmological solutions having
direct physical interest. Properties of these solutions are studied below in sections~3--5 in
dependence on initial conditions at a bounce, restrictions on parameter $\alpha$ and parameters of
scalar field potential. Note that, if the value of $\alpha^{-1}$ is much less than the Planckian
energy density, equations (13)--(14) with scalar field potentials applying in inflationary
cosmology lead also to oscillating solutions studied in Ref. \cite{mc7}.

\section{Some properties of inflationary cosmological solutions}

Let us introduce 4-dimensional phase space $P$ with the axes ($\tilphi$, $\tilphi'$, $tilH$,
$\tilde{\rho}_{\mathrm{m}}$). Then some trajectory in space $P$ corresponds to any solution of
GCFE. Trajectories of cosmological solutions are situated in area of $P$, where $\tilZ\geq 0$ and,
hence, are limited by the bounds $L$ defined as
\begin{equation} \label{e13}
\tilZ=0\qquad \textrm{or}\qquad \tilphi'=\pm\sqrt{1+4\tilV\,}.
\end{equation}
According to equation (13) values of the Hubble parameter  are equal to
\begin{equation}
\label{e15}
\tilH_{\pm}=
    \frac{\dst -3\frac{\pat \tilV}{\pat \tilphi}\, \tilphi' \pm \sqrt{\tilde{D}}}
    {\dst \tilZ+3\tilphi'^2},
\end{equation}
where
\begin{equation}
\label{e16}
\tilde{D}=
    \frac{8\pi}{3}\,
    \left[\tilrhm+\frac{1}{2}\tilphi'^2+\tilV +\frac{1}{4}
        \left(4\tilV-\tilphi'^2\right)^2
    \right]\,\tilZ-\frac{k}{\tilR^2}\tilZ^2.
\end{equation}
Two values of the Hubble parameter $\tilH_{+}$ and $\tilH_{-}$ correspond to two different
solutions named as $\tilH_{\pm}$-solutions. In points of bounds $L$ we have $\tilde{D}=0$,
$\tilH_{+}=\tilH_{-}$ and the Hubble parameter is determined by
\begin{equation}
\label{e14}
\tilH_{L}=-\frac{1}{\tilphi'}\,\frac{\pat \tilV}{\pat \tilphi}\, .
\end{equation}
Now let us consider extremal $H_0$-surfaces in the space $P$, in points of which $\tilH_{+}=0$ or
$\tilH_{-}=0$. Denoting the values of quantities on $H_0$-surfaces by means of index ``0'', we
obtain from (13) the following equation for $H_0$-surfaces, which are in fact in 3-dimensional
hyperspace $\tilH=0$ of the phase space $P$:
\begin{eqnarray}
\label{e17}
\frac{k}{\tilR_0^2}\,\tilZ_0^2 +9{\left(\frac{\pat \tilV}{\pat\tilphi}\right)\!}_0^2\tilphi_0'^2
=\frac{8\pi}{3}\,
 \left[
   \tilrhmz+\frac{1}{2}\tilphi_0'^2+\tilV_0
   +\frac{1}{4}\left(4\tilV_0-\tilphi_0'^2\right)^2
 \right]
\,\tilZ_0,
\end{eqnarray}
where $\tilZ_0=1+4\tilV_0-\tilphi_0'^2$. For given scalar field potential $\tilV(\tilphi)$ this
equation determines $H_0$-surfaces for flat models ($k=0$), and in the case of closed and open
models equation (19) gives parametric families of $H_0$-surfaces with parameter $\tilR_0$.
According to equation (14) the time derivative of the Hubble parameter $\tilH'_0$ on $H_0$-surfaces
is
\begin{eqnarray}
\label{e18}\fl 
\tilH'_0=
\left\{
    8\pi\left[
        \tilV_0+\frac{1}{3}\tilrhmz
        +\frac{1}{4}\left(4\tilV_0-\tilphi_0^2\right)^2
    \right]
    \right.\nonumber\\
    \left.
    -3\left[
        \left(\frac{\pat^2\tilV}{\pat\tilphi^2}\right)_0\tilphi_0'^2
        -\left(\frac{\pat\tilV}{\pat\tilphi}\right)_0^2
    \right]
    -\frac{2k}{\tilR_0^2}\tilZ_0
\right\}\left(\tilZ_0+3\tilphi_0'^2\right)^{-1}\, .
\end{eqnarray}
If $\tilH'_0>0$, we have in corresponding points of $H_0$-surfaces the transition from compression
to expansion (a bounce), and in the case $\tilH'_0<0$ the transition from expansion to compression
takes place. Equations (19)--(20) permit to find specific points of solutions, where $\tilH=0$ and
$\tilH'=0$ simultaneously.

To analyze solutions of GCFE we have to study $H_0$-surfaces defined by (19) for given scalar field
potential. We will use below the simplest quadratic potential in the form $V=\frac{1}{2}m^2\phi^2$
or $\tilV=\frac{1}{2}\tilm^2\tilphi^2$, where $\tilm=m\sqrt{\frac{\alpha}{G}}$. In this case
$H_0$-surfaces depend on parameter $\tilm$ and the value of scale factor $\tilR_0$. In the case of
$\tilrhm=0$ equation (19) for $H_0$-surfaces includes only variables $\tilphi_0$ and $\tilphi'_0$,
and the analysis of $H_0$-surfaces as well as bounds $L$ is reduced to analysis of corresponding
curves on the plane ($\tilphi$, $\tilphi'$). Numerical analysis will be given below in the
Planckian system of units with $\hbar=c=G=1$. Note, that admissible values of dimensionless
potential $\tilV=\alpha V$ can be essentially greater than 1, if the scale of extremely high energy
densities defined by $\alpha^{-1}$ is much less then the Planckian one. Hence, values of parameter
$\tilm$ can be large. Below the analysis is given for various types of cosmological inflationary
models.

\section{Numerical analysis of flat inflationary models}

\begin{figure}[htb!]
\begin{minipage}{0.48\textwidth}\centering{
 \includegraphics[width=\linewidth]{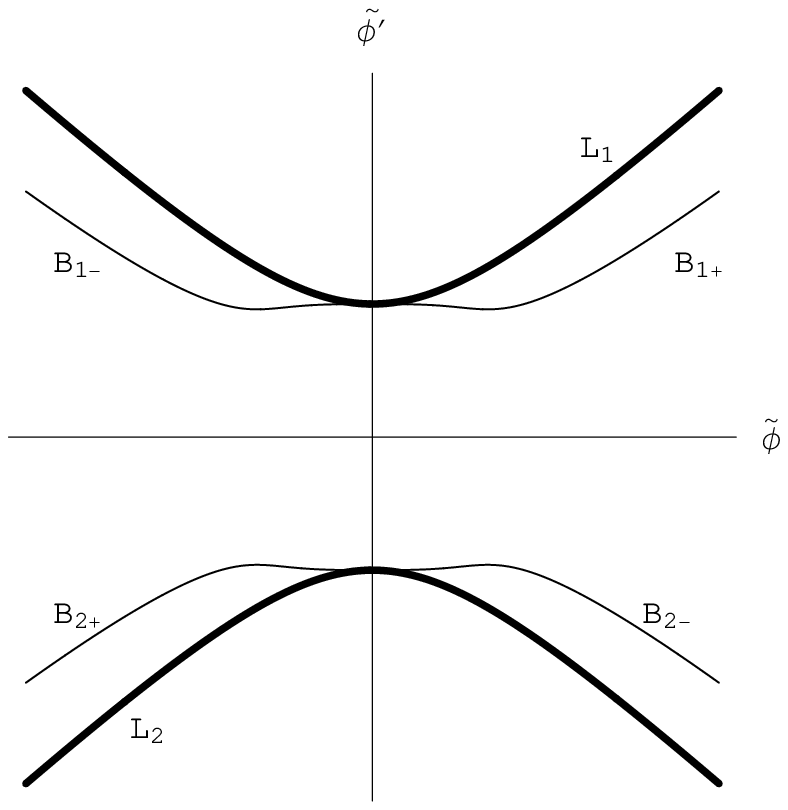}
}
\end{minipage}\, \hfill\,
\begin{minipage}{0.48\textwidth}\centering{
 \includegraphics[width=\linewidth]{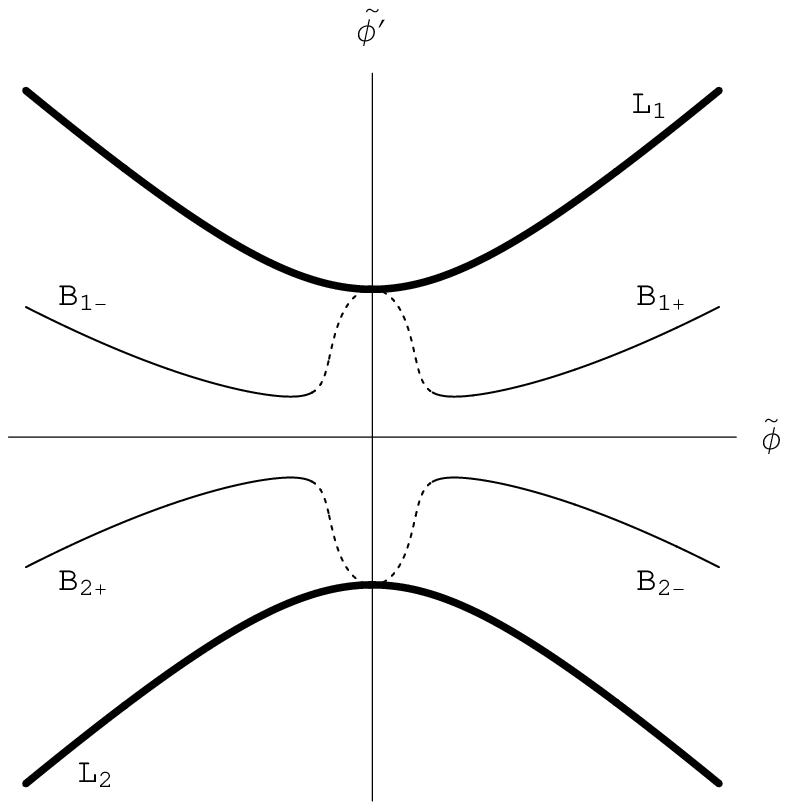}
}
\end{minipage}
\caption{\label{figm2}Bound $L_{1,2}$-curves and bounce curves $B_{1,2}$ for flat models in the
case of potential $\tilV=\frac{1}{2}\tilm^2\tilphi^2$. Dotted curves correspond to negative values
of $\tilH'_0$.}
\end{figure}

In the case of flat models ($k=0$) $H_0$-curves $B_{1,2}$ are presented in Fig.~1 together with
bounds $L_{1,2}$ having the form of two hyperbolas $\tilphi'= \pm \sqrt{1+2\tilm^2\tilphi^2}$. Each
of two curves $B_{1,2}$ contains two parts corresponding to vanishing of $\tilH_+$ or $\tilH_-$ and
denoting by ($B_{1+}$, $B_{2+}$) and ($B_{1-}$, $B_{2-}$) respectively \cite{mc1}. The parts
($B_{1+}$, $B_{1-}$) and ($B_{2+}$, $B_{2-}$) have the common points ($\tilphi=0,\tilphi'=\pm 1$)
with the bounds $L_1$ and $L_2$ on the axis $\tilphi'$. From equations (19) -- (20) it is possible
to show, that if
$$
\tilm<2\sqrt{\frac{\pi}{42}}\sqrt{-3-47\left(1009+84
   \sqrt{159}\right)^{-1/3}+\left(1009+84 \sqrt{159}\right)^{1/3}}\approx 1{.}346,
$$
than the value of $\tilH'_0$ on $H_0$-curves is positive (this means that $\tilH'_{+0}>0$ in points
of $B_{1+}$ and $B_{2+}$, and $\tilH'_{-0}>0$ in points of $B_{1-}$ and $B_{2-}$). In this case
$B_{1,2}$-curves are ``bounce curves'' \cite{mc1}, in points of which the transition from
compression to expansion takes place. If $\tilm>1{.}346$ (that corresponds to large values of
$\alpha$), the regions with $\tilH'_0<0$ appear on $H_0$-curves near the axis $\tilphi'$ for
sufficiently small values of $|\tilphi|$ (see Fig.~\ref{figm2}). In this case, besides regular
cosmological solutions, the GCFE have scalar field oscillating solutions \cite{mc7}.

\begin{figure}[htb!]
\begin{minipage}{0.48\textwidth}\centering{
 \includegraphics[width=\linewidth]{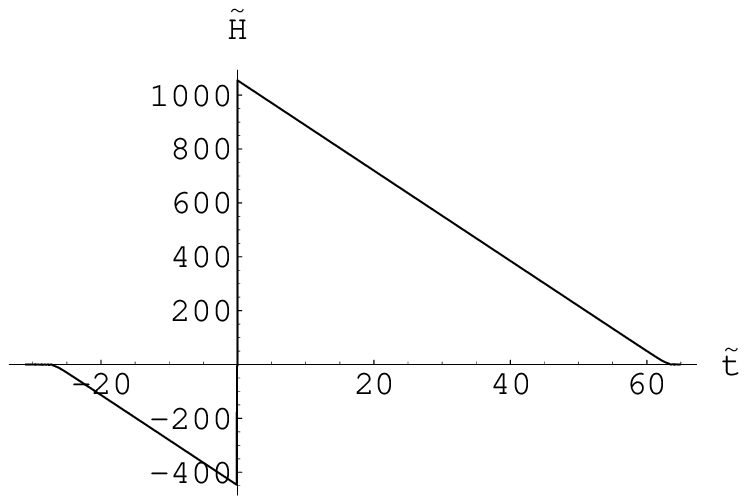}
}
\end{minipage}\, \hfill\,
\begin{minipage}{0.48\textwidth}\centering{
 \includegraphics[width=\linewidth]{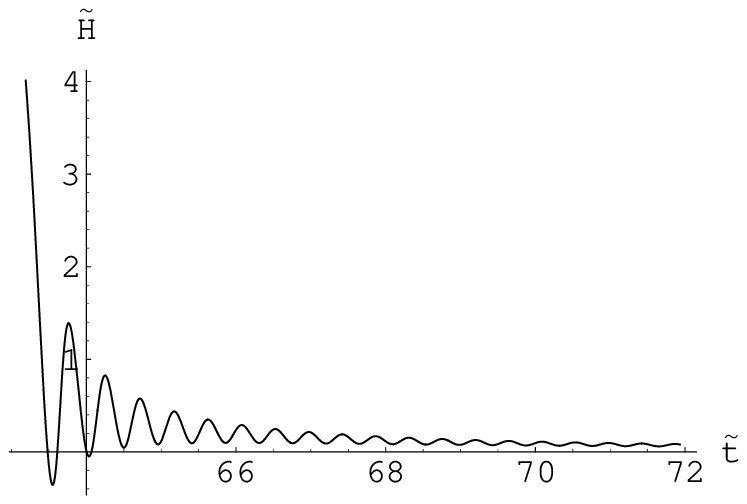}
}
\end{minipage}
\caption{\label{figm2a}Compression and inflationary stages; postinflationary stage.}
\end{figure}

\begin{figure}[htb!]
\,\hfill
\begin{minipage}{0.48\textwidth}\centering{
 \includegraphics[width=\linewidth]{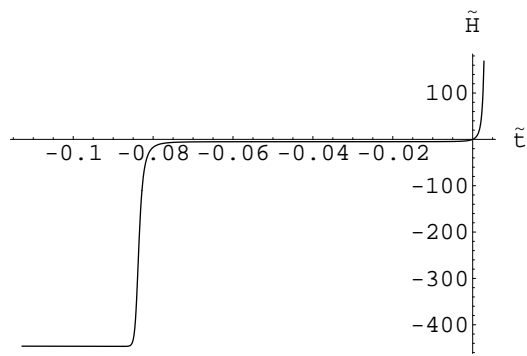}
}
\end{minipage}\, \hfill\,
\caption{\label{figm2b}Transition stage}
\end{figure}

Now let us analyze properties of cosmological solutions for flat models. Each such solution,
generally speaking, includes the quasi-de-Sitter compression and inflationary stages, stage after
inflation (Fig.~\ref{figm2a}) and the stage of transition from compression to expansion
(Fig.~\ref{figm2b}). Properties of cosmological solutions --- duration of different stages of
evolution, characteristics of scalar field oscillations and behaviour of the Hubble parameter after
inflation and before the compression stage are studied in dependence on initial conditions at a
bounce and the value of $\tilm$. Like to GR \cite{mc9}, the initial value of $\tilphi_0$ must be
greater than 1 ($\tilphi_0>1$) to appear inflationary stage.

\begin{figure}[htb!]
{}\hfill
\begin{minipage}{0.48\textwidth}\centering{
 \includegraphics[width=\linewidth]{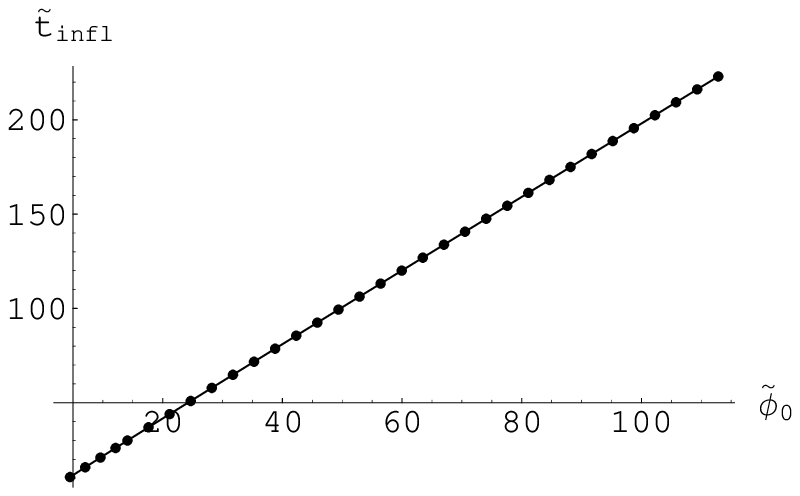}
}
\end{minipage}\, \hfill\,
\begin{minipage}{0.48\textwidth}\centering{
 \includegraphics[width=\linewidth]{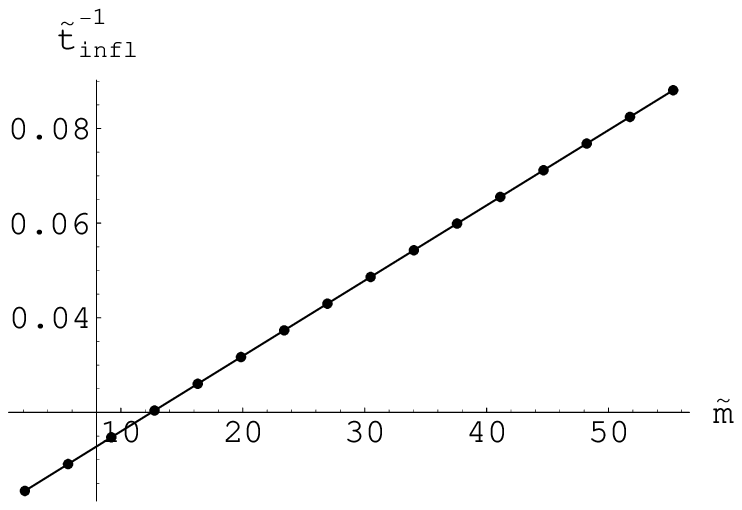}
}
\end{minipage}\hfill \,
\caption{\label{figm3}Duration of inflationary stage in dependence on values $\tilphi_0$ and
$\tilm$.}
\end{figure}

Because GCFE lead to slow-rolling regime which coincide practically with that of GR, properties of
inflationary stage of our models are the same as in GR. Namely, the duration of inflationary stage
$t_{\mathrm{infl}}$ increases linearly by increasing of $|\tilphi_0|$ and decreases inversely
proportional by increasing of $\tilm$ (Fig.~\ref{figm3}).

Because of $\tilphi'_0\neq 0$, cosmological solutions are asymmetric in the time with respect to
the bounce point. In particular, the duration of quasi-de-Sitter stage of compression is not equal
to duration of inflationary stage. In the case of bounce in points of curves $B_{1+}$ and $B_{2+}$
the duration of inflationary stage is greater than duration of quasi-de-Sitter compression stage.
If the bounce takes place in points of bounce curves $B_{1-}$ and $B_{2-}$ the situation is
opposite.

\begin{figure}[htb!]
\begin{minipage}{0.48\textwidth}\centering{
 \includegraphics[width=\linewidth]{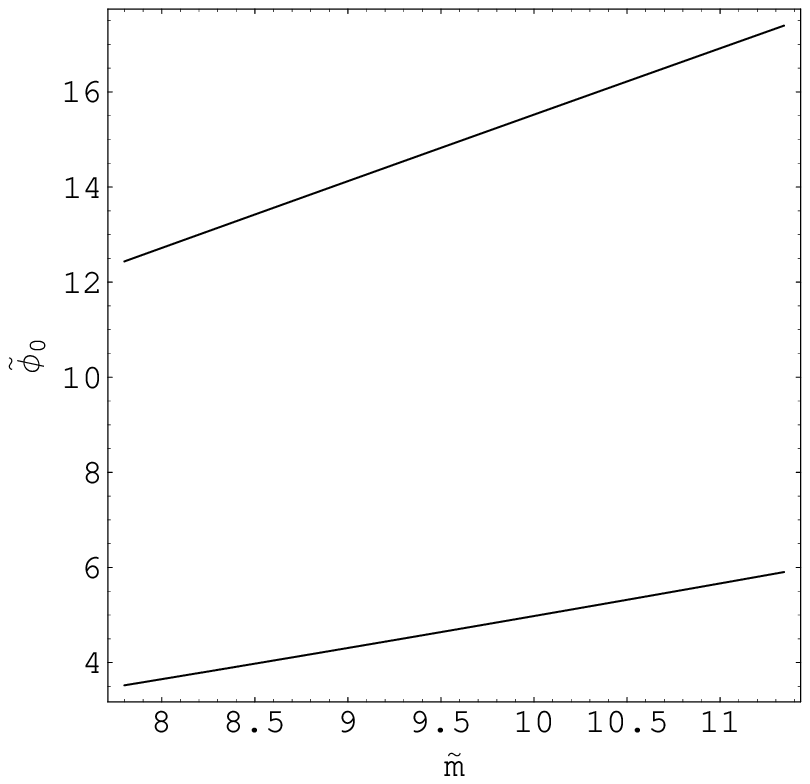}
}
\end{minipage}\, \hfill\,
\begin{minipage}{0.48\textwidth}\centering{
 \includegraphics[width=\linewidth]{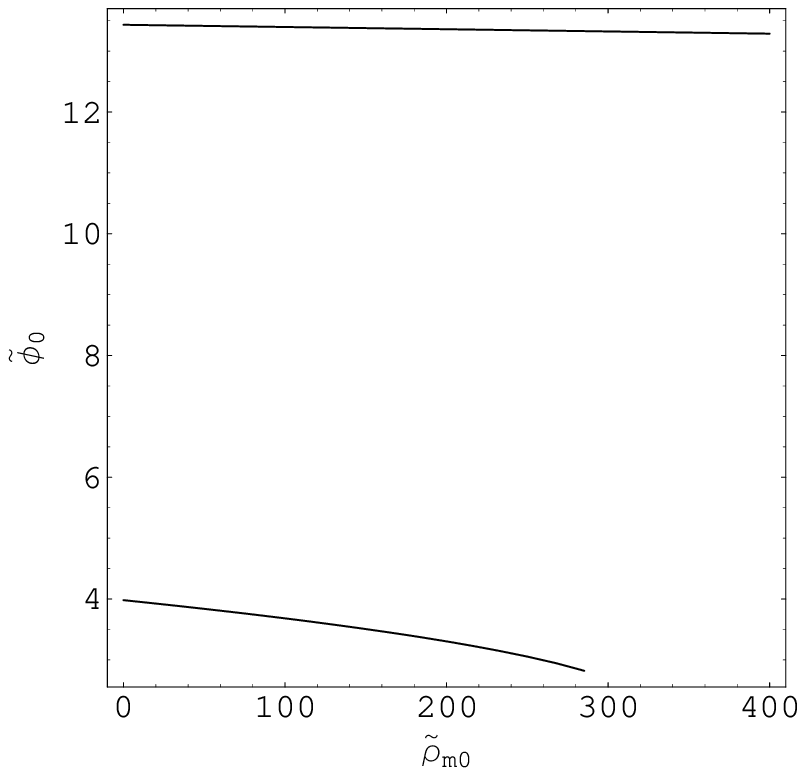}
}
\end{minipage}
\caption{\label{figm4} The upper and lower bounds of interval for admissible values of $\tilphi_0$
for solutions without quasi-de-Sitter compression stage.}
\end{figure}

For given value of $\tilm$ inflationary models without quasi-de-Sitter compression stage
($\tilrhm=0$) can be realized for certain interval of $\tilphi_0$. For example, at $\tilm=1{.}2$
corresponding interval of $\tilphi_0$ is ($28$, $100$), then the growth of the scale factor $R(t)$
is in interval ($e^{130}$, $e^{1450}$). By increasing of $\tilm$ the upper and lower bounds of
interval for $\tilphi_0$ of discussed solutions increase linearly (Fig.~\ref{figm4}a). In the case
of presence of radiation ($\tilrhm\neq 0$) the lower bound of interval of $\tilphi_0$ for solutions
without quasi-de-Sitter stage of compression decreases essentially and upper bound decreases slowly
(Fig.~\ref{figm4}b).

\begin{figure}[htb!]
\begin{minipage}{0.48\textwidth}\centering{
 \includegraphics[width=\linewidth]{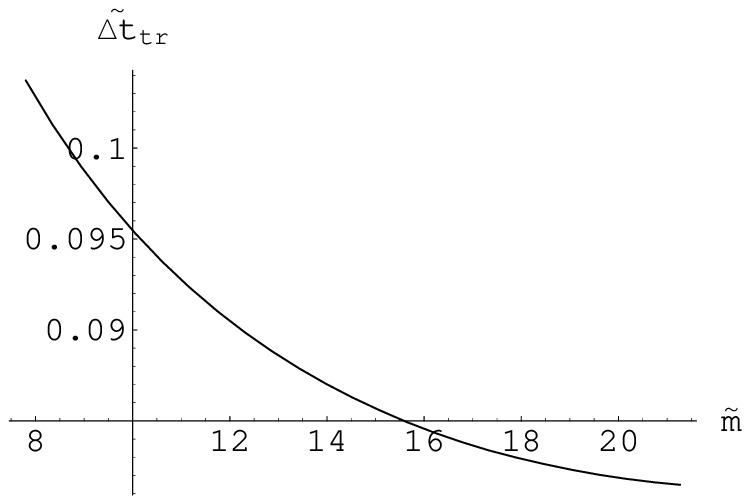}
}
\end{minipage}\, \hfill\,
\begin{minipage}{0.48\textwidth}\centering{
 \includegraphics[width=\linewidth]{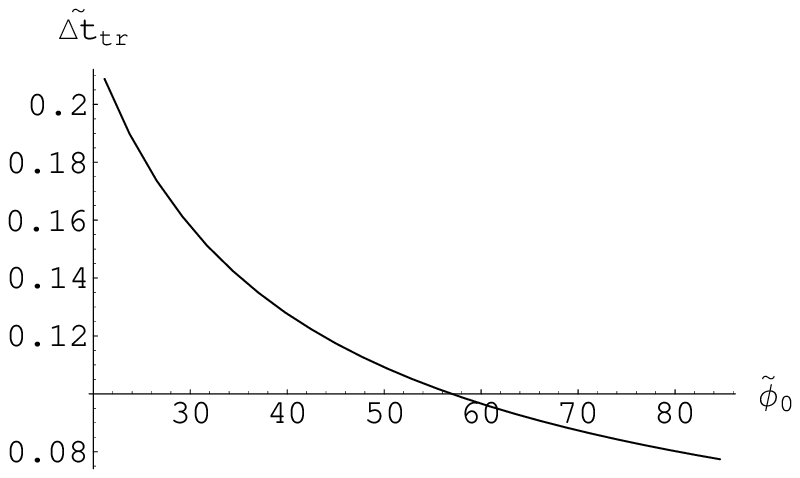}
}
\end{minipage}
\caption{\label{figm5}Duration of transition stage in dependence on values of $\tilm$ and
$\tilphi_0$.}
\end{figure}

\begin{figure}[htb!]
\,\hfill\,
\begin{minipage}{0.6\textwidth}\centering{
 \includegraphics[width=\linewidth]{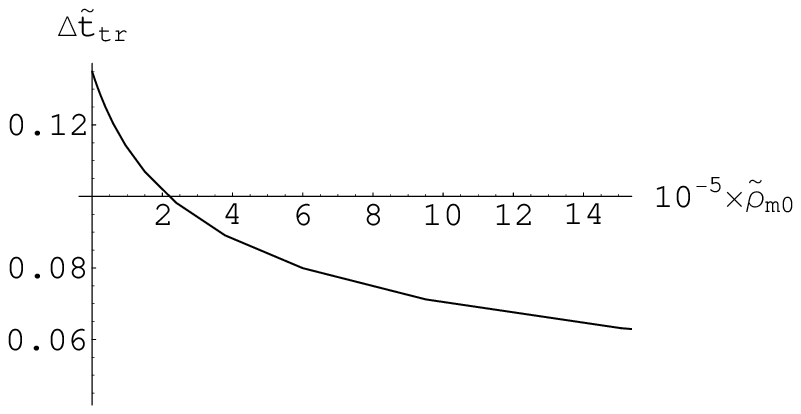}
}
\end{minipage}\, \hfill\,
\caption{\label{figm6}Duration of transition stage in dependence on $\tilrhmz$.}
\end{figure}

The most principal feature of our cosmological models is the presence of regular transition stage
from compression to expansion. The behaviour of scalar field $\tilphi$ and the Hubble parameter
$\tilH$ at transition stage is essentially nonlinear and complicated. The duration of transition
$\Delta\tilt_{\mathrm{tr}}$ stage decreases by increasing of each of parameters $\tilm$,
$\tilphi_0$ and $\tilrhmz$ (Fig.~\ref{figm5} -- Fig.~\ref{figm6}).

\begin{figure}[htb!]
\begin{minipage}{0.48\textwidth}\centering{
 \includegraphics[width=\linewidth]{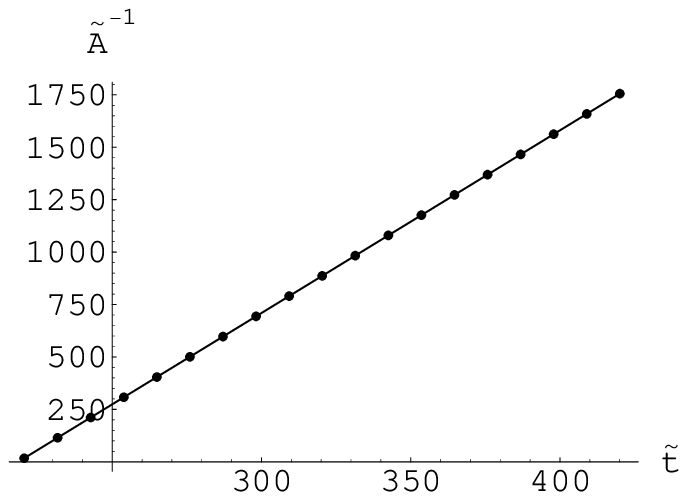}
}
\end{minipage}\, \hfill\,
\begin{minipage}{0.48\textwidth}\centering{
 \includegraphics[width=\linewidth]{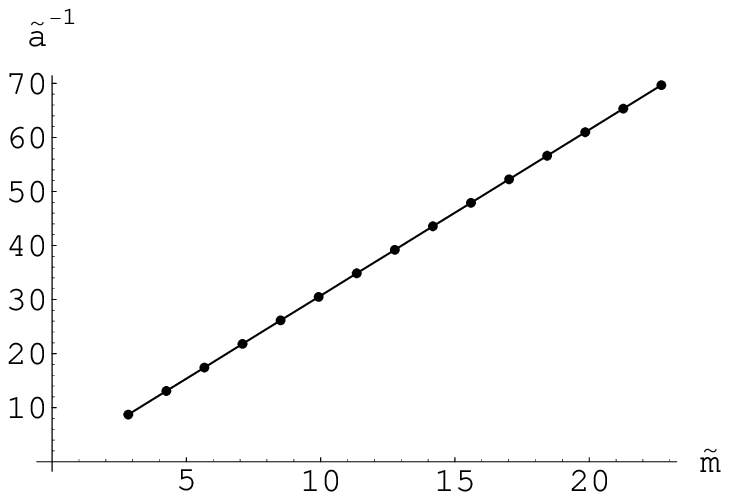}
}
\end{minipage}
\caption{\label{figm7}Dependence of amplitude of $\tilphi$-oscillations on the time $\tilt$ and on
$\tilm$ at $\tilm<30$.}
\end{figure}

\begin{figure}[htb!]
\begin{minipage}{0.48\textwidth}\centering{
 \includegraphics[width=\linewidth]{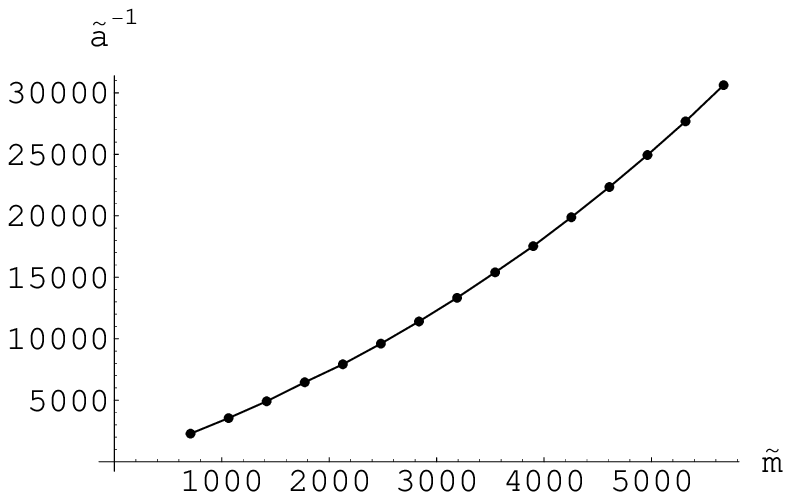}
 \caption{\label{figm8} Dependence of $\tilphi$-oscillations amplitude at large values of $\tilm$.}
}
\end{minipage}
\,\hfill\,
\begin{minipage}{0.48\textwidth}\centering{
 \includegraphics[width=\linewidth]{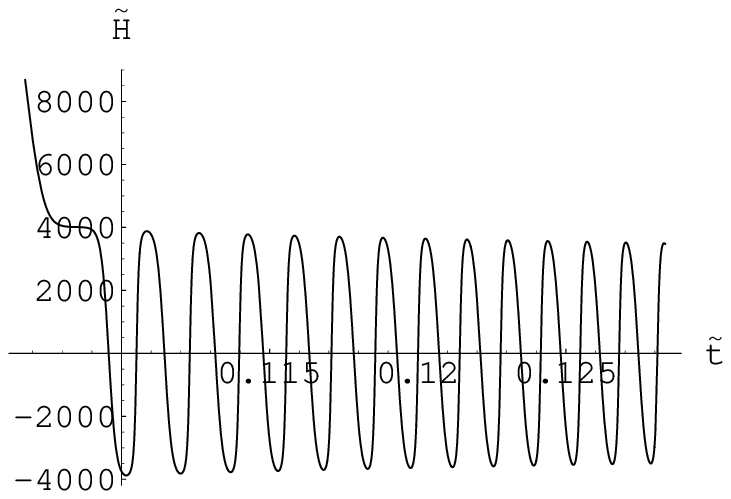}
 \caption{\label{figm8a} $\tilH$-oscillations after inflation at large values of $\tilm$.}
 }
\end{minipage}
\end{figure}

After inflation stage the scalar field oscillates near its minimum. These oscillations don't depend
on initial values of $\tilphi_0$ and $\tilrhmz$, but essentially depend on parameter $\tilm$. Like
to GR frequency and amplitude of scalar field oscillations decreases with the time, but
quantitatively such dependence is different. Note, that energy density $\tilrhm$ is negligibly
small at the end of inflationary stage, even if $\tilrhmz\neq0$. Application of Krylov-Bogoliubov
method \cite{kr} of solutions of the system of differential equations (10) and (14) with
$\tilrhm=0$, after some tedious calculations, leads to the following asymptotics for scalar field
$\tilphi(t)$ and the scale factor $\tilR(t)$ at late time
\begin{equation}
\label{phiasymp}
\tilphi(\tilt)\approx \frac{1}{\sqrt{3\pi}\,\tilm\tilt}\,
    \cos \left(\frac{3\tilm^2+2\pi}{8\pi\tilm\tilt}+\tilm\tilt\right),
\end{equation}
\begin{equation}
\label{Rasymp}
\tilR(\tilt)\approx \sqrt[3]{16\pi}\,\tilR_0
\tilt^{2/3}-\frac{\sqrt[3]{16\pi}\,\tilR_0}{\tilm^2\tilt^{4/3}}\,
    \cos{\left(\frac{27 \tilm^4+36 \pi  \tilm^2-20 \pi ^2}{12\pi\tilm\left(3\tilm^2+2
   \pi\right)\tilt}+2\tilm\tilt\right)}.
\end{equation}
Equation (\ref{phiasymp}) shows, that the amplitude of scalar field oscillations $\tilde{A}$
decreases with time as follows $\tilde{A}=\tilde{a}(\tilm)\tilt^{-1}$, where
$\tilde{a}=4\tilm^{-1}/\sqrt{3}$. Numerical analysis confirms relations
(\ref{phiasymp})--(\ref{Rasymp}) at sufficiently small values of $\tilm$ ($\tilm\lesssim 30$), in
particular, the amplitude of oscillations decreases inverse proportional to time (at
$\tilm=\mathrm{const}$) and inverse proportional to $\tilm$ (Fig.~\ref{figm7}).\footnote{According
to numerical calculation, constant of proportionality at Fig.~\ref{figm7}b is equal to 3{.}070,
which is close to $\sqrt{3\pi}\approx 3{.}069$ as Krylov-Bogoliubov method requires.} But at large
values of $\tilm$ ($\tilm\sim 1000$) conclusions obtained by means of Krylov-Bogoliubov method
differ from numerical results. In particular, the inverse proportional dependence between the
amplitude of scalar field oscillations and the value of $\tilm$ is not valid (Fig.~\ref{figm8}).
This is connected with the fact, that in the case of large values of $\tilm$ (i.e. large values of
$\alpha$) the Hubble parameter, unlike to GR, oscillates near the value $\tilH=0$
(Fig.~\ref{figm8a}). The frequency of $\tilH$-oscillations is two times greater than the frequency
of $\tilphi$-oscillations.  The value of $\tilH$ becomes positive with the time and tends
asymptotically to zero like to GR. Note that, there are similar $\tilphi$-oscillations and
$\tilH$-oscillations before the compression stage also.

\section{Numerical analysis of open and closed models}

In the case of open models $H_0$-curves are similar to that of flat models and the family of
$H_0$-curves is situated on the plan ($\tilphi$,$\tilphi'$) between curves $L_1$ and $B_1$, and
also between curves $L_2$ and $B_2$. At $R_0\to 0$ $H_0$-curves of open models tend to bounds
$L_{1,2}\,$, and at $R_0\to\infty$ they approach to curves $B_{1,2}$ (see Fig.~1). In accordance
with expression (21) for $\tilH'_0$ the minimum value of $\tilm$, by which negative values of
$\tilH'_0$ appear for open models, depends on $\tilR_0$ and it is smaller then that for flat
models. Properties of cosmological solutions for open models are similar to that of flat models.
There are area of $\tilH'<0$ on $H_0$-curves, if $\tilm>2\sqrt{\pi/6}\approx 1{.}45$  and the value
of $\tilR_0$ is arbitrary or $\tilm<2\sqrt{\pi/6}$ and
$$\tilR<\sqrt{\dst\frac{3}{4\pi}}\sqrt{\dst\frac{1+24\tilm^2}{1+30\tilm^2+216\tilm^4}}\;.$$
Note that in the case of open models there are also solutions with $\tilZ<0$, which are separated
from cosmological solutions by bounds $L_{1}$ and $L_{2}$. These solutions are hardly ever have
physical sense.

\begin{figure}[hbt]
\centering{
 \begin{minipage}{0.6\textwidth}
 \includegraphics[width=\textwidth]{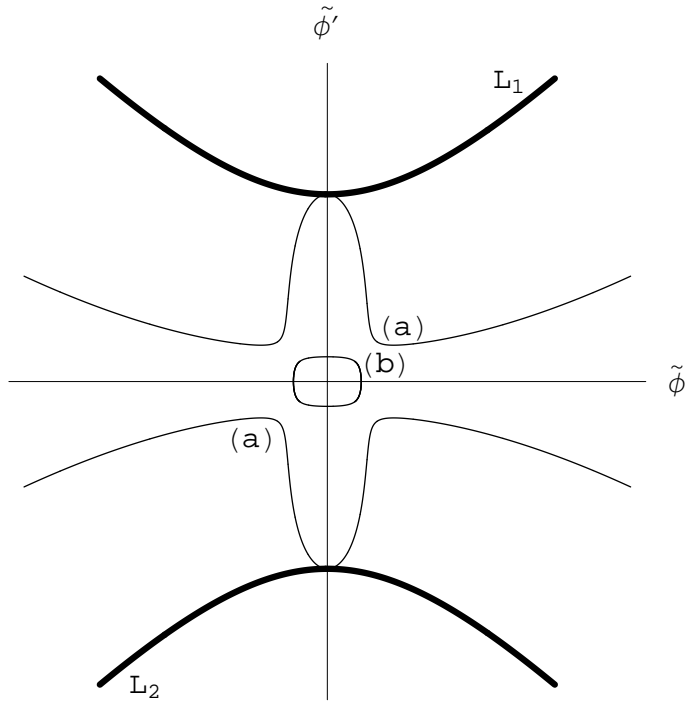}
 \label{mnfig2}
 \end{minipage}}
 \caption{$\tilH_0$-curves for closed models at large values $\tilR_0$.}
 \end{figure}

\begin{figure}[htb]
\begin{minipage}{0.48\textwidth}
 \includegraphics[width=\textwidth]{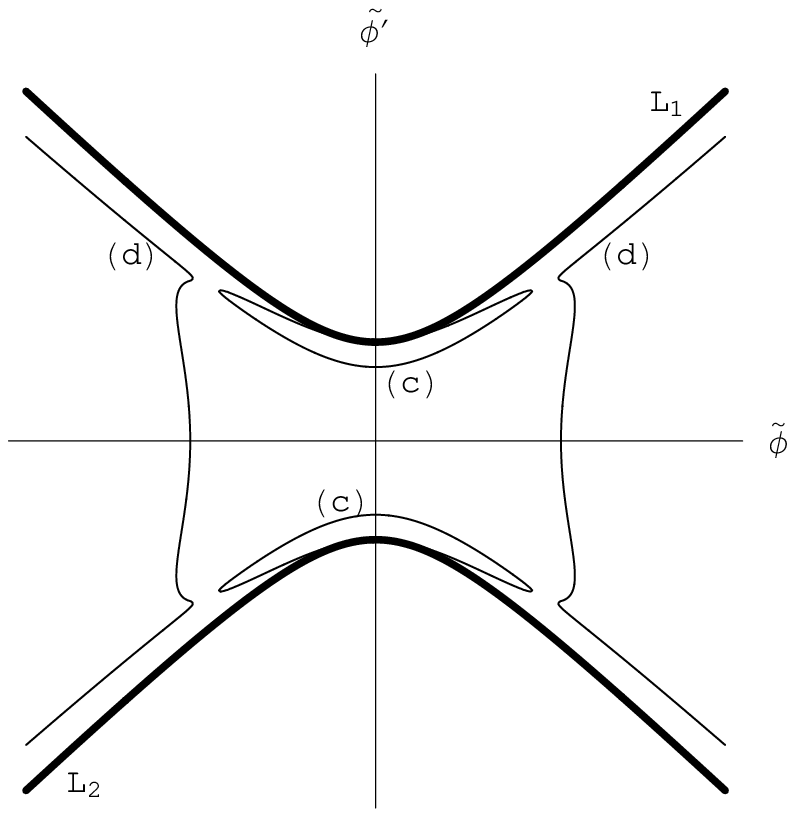}
 \end{minipage}
 \hfill
 \begin{minipage}{0.48\textwidth}
 \includegraphics[width=\textwidth]{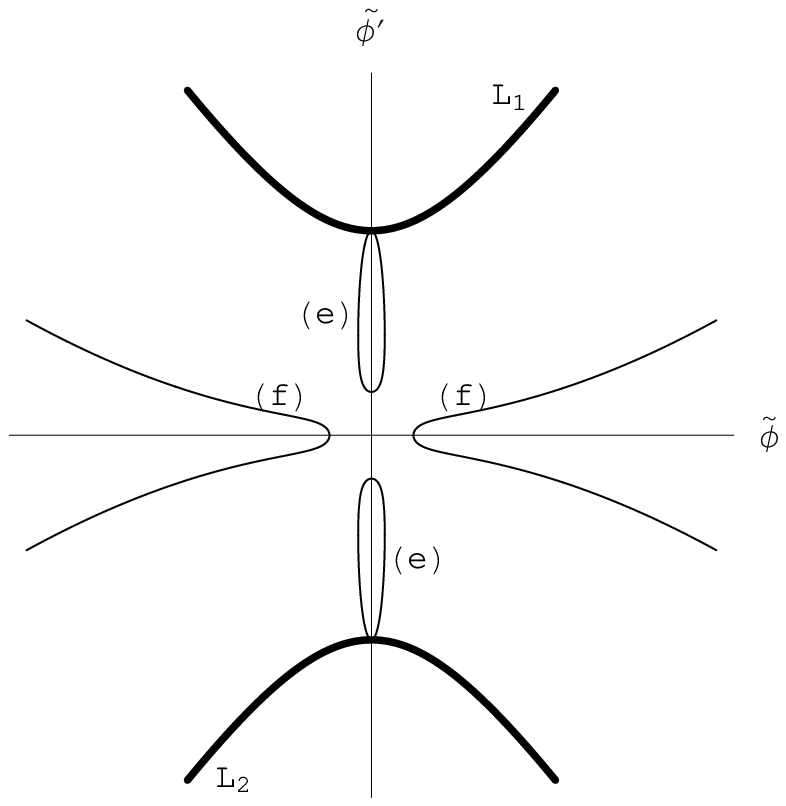}
 \end{minipage}
 \caption{$\tilH_0$-curves for closed models at small values of $\tilR_0$.}\label{mnfig3}
 \end{figure}

The form of $H_0$-curves for closed models ($k=+1$) differs essentially from that for flat and open
models, and it depends on the value of $\tilR_0$ for given value of $\tilm$. For sufficiently large
values of $\tilR_0$ $H_0$-curves for closed models include besides two curves similar to that of
flat model also one closed curve with center in origin of coordinates (Fig.~\ref{mnfig2}). If
$\tilm<1{.}34$ and the value of $\tilR_0$ is sufficiently large, the value of $\tilH'_0$ is
positive on $H_0$-curves $(a)$ and on the part of $H_0$-curve $(b)$. In the case $\tilm>1{.}34$ the
region with $\tilH'_0<0$ appears on $H_0$-curves $(a)$ near the axes $\tilphi'$. By decreasing of
$\tilR_0$ the $H_0$-curves of closed models are deformed and transformed into $H_0$-curves
presented in Fig.~\ref{mnfig3}. If $\tilm<1{.}4$ $H_0$-curves $(c)$ contain regions with
$\tilH'_0>0$ and regions with $\tilH'_0<0$; small region with $\tilH'_0<0$ can be present on
$H_0$-curves $(d)$ although  on the greatest part of $H_0$-curves $(d)$ we have $\tilH'_0>0$. In
the case $\tilm>1{.}4$ and $\tilR_0<\tilR_{01}$ (for example, $\tilR_{01}=1{.}5$ for $\tilm=2{.}1$,
$\tilR_{01}=2{.}5$ for $\tilm=3{.}5$)\footnote{The value $\tilR_{01}^2$ is the real root of cubic
equation, which can be obtained from equation (19) $\pi (-243 \tilm^6-306 \pi  \tilm^4-408 \pi ^2
\tilm^2+32 \pi ^3) \tilR_{01}^6+3 (6561 \tilm^8+7776 \pi  \tilm^6+12960 \pi ^2 \tilm^4-5376 \pi ^3
\tilm^2+256 \pi^4) \tilR_{01}^4-6912 \tilm^2 \pi  (81 \tilm^4-144 \pi  \tilm^2+16 \pi ^2)
\tilR_{01}^2+3981312 \tilm^4 \pi^2=0$} we have $\tilH'_0>0$ on the curves $(f)$ and $\tilH'_0<0$ on
the $H_0$-curves $(e)$. Like to flat and open models oscillating solutions exist for large values
$\tilm$ ($\alpha$). Unlike to flat and open models, in the case of closed models there are
cosmological solutions symmetric in the time with respect to a bounce ($\tilphi'_0=0$). Note, that
cosmological solutions have properties similar to flat models, if the value of $\tilR_0$ at a
bounce is sufficiently large.


\section{Conclusion}

Numerical analysis of inflationary cosmological models curried out in the framework of gauge
theories of gravitation confirms obtained earlier conclusion about their regular character. The
difference in behaviour of cosmological models at the end of inflationary stage in comparison with
GR (in the case of large values of $\alpha$) can be interesting, if we take into account recent
results concerning the anisotropy of cosmic microwave background \cite{mc10}. This means that
further investigations of perturbations in discussed inflationary models is of direct physical
interest.


\begin{thebibliography}{99}
\bibitem{mc1} Minkevich A V 2005 ({\it Preprint} gr-qc/0506140), submitted to {\it Gravitation \& Cosmology}
\bibitem{mc2} Minkevich A V 2005 {\it Int. J. Mod. Phys.} A {\bf 20} 2436--42.
\bibitem{mca3} Hehl F W 1980 {\it In: Cosmology and Gravitation} (New York: Plenum Press)
\bibitem{mca4} Hehl F W, McGrea G D, Mielke E W  and Neeman Y 1995 {\it Phys. Rep.}
{\bf 258} 1
\bibitem{mc3} Minkevich A V 2005 Gravitational repulsion effect at extreme conditions ({\it Preprint}
gr-qc/0512123)
\bibitem{mc4} Minkevich A V 1980 {\it Vestsi Akad. Nauk BSSR.
Ser. fiz.-mat.}, no.~2 87; 1980 {\it Phys. Lett.} A {\bf 80} 232
\bibitem{mc5} Minkevich A V 1993 {\it Dokl. Akad. Nauk Belarus.} {\bf 37} 33
\bibitem{mc6} Minkevich A V and Garkun A S 2000 {\it Class. Quantum
Grav.} {\bf 17} 3045
\bibitem{mc7} Minkevich A V, Garkun A S and Vasilevski Yu G 2004
{\it Nonlinear Phenomena in Complex Systems} {\bf 7} \No 1 78 ({\it Preprint} gr-qc/0310060)
\bibitem{mc9} Linde A D 1990 {\it Particle Physics and Inflationary Cosmology} (Harwood: Chur,
Switzerland)
\bibitem{kr} Bogoliubov N N and Mitropolsky Y A 1961 {\it Asymptotic Methods in the Theory of Non-Linear
Oscillations} (New York: Gordon and Breach Science Publishers)
\bibitem{mc10} Copi C J, Huterer D, Schwarz D J and Starkman G D 2005 On the large-angle anomalies of the microwave
sky ({\it Preprint} astro-ph/0508047)
\end{thebibliography}
\end{document}